# Unmasking Falsehoods in Reviews: An Exploration of NLP Techniques

Anusuya Baby Hari Krishnan

College of IT, United Arab Emirates University, UAE

## Abstract

In the contemporary digital landscape, online reviews have become an indispensable tool for promoting products and services across various businesses. Marketers, advertisers, and online businesses have found incentives to create deceptive positive reviews for their products and negative reviews for their competitors' offerings. As a result, the writing of deceptive reviews has become an unavoidable practice for businesses seeking to promote themselves or undermine their rivals. Detecting such deceptive reviews has become an intense and ongoing area of research. This research paper proposes a machine learning model to identify deceptive reviews, with a particular focus on restaurants. This study delves into the performance of numerous experiments conducted on a dataset of restaurant reviews known as the Deceptive Opinion Spam Corpus. To accomplish this, an n-gram model and max features are developed to effectively identify deceptive content, particularly focusing on fake reviews. A benchmark study is undertaken to explore the performance of two different feature extraction techniques, which are then coupled with five distinct machine learning classification algorithms. The experimental results reveal that the passive aggressive classifier stands out among the various algorithms, showcasing the highest accuracy not only in text classification but also in identifying fake reviews. Moreover, the research delves into data augmentation and implements various deep learning techniques to further enhance the process of detecting deceptive reviews. The findings shed light on the efficacy of the proposed machine learning approach and offer valuable insights into dealing with deceptive reviews in the realm of online businesses.

**Keywords:** data mining, fake reviews, ecommerce industry, opinion spams, text classification

## 1. Introduction:

In this era of technology, many people can post their opinions on several online websites. These online reviews play an important role for the organizations and for future customers, who get an idea about products or services before making a decision or selection. Additionally, many search engine optimization processes will give high rank to businesses when they have more positive reviews.

Generally people can post their opinions about particular products on ecommerce platforms like Amazon, Volusion, Shopify, BigCommerce, Magento, WooCommerce, Wix, and Big Cartel etc. So online reviews make a huge impact on people across e-commerce industries, where personal reviews on products are considered to be useful to make a decision whether to purchase a product or not. For example, if customer wants to do online shopping in any ecommerce platforms, they definitely read online reviews on the opinions of other customers. Depending on the feedback of the online reviews, customer gets deep insight into the products they are planning to buy. So a new customer makes their decisions of whether to purchase the products or not by analyzing the particular ecommerce websites to see the other people's opinions on those products. If they came across a lot of positive feedback from the online reviews about a particular product, they probably go on to buy the product. Thus, online reviews became very important sources of information for customers.

But some people misdirect others by posting deceptive reviews to promote or degrade the reputation of some other particular products as per desire. Since reviews have some authentic feedbacks about positive or negative services, many businesses manipulate customers' decisions by giving inauthentic content about the product. This is considered as deceptive and such reviews are called as fake review. These people who give misleading opinions are called opinion spammers.

In our research paper, we explore the opinion spam namely, fake reviews. We propose an integrated approach for the detection of deceptive content, applicable for detecting fake review. We use lemmatization for data cleaning and we analyse a detection model that combines both n-gram and max features. We also investigate two different features extraction techniques and five different machine learning classification techniques.

The proposed model is evaluated using Ott et al.'s dataset, involving truthful and fake review content. The experimental evaluation indicates that our model outperforms existing work using the same dataset. We achieved an accuracy of 92.5% accuracy which is slightly higher than the existing work which is 90% achieved by Ahmed et al. on the same dataset.

The rest of the paper tells us as follows. Section 2 summarizes the review of related works. Section 3 introduces the background and the details of the proposed machine learning approach. Section 4 presents two experiments conducted to evaluate the accuracy of our model in identifying deceptive reviews. Section 5 concludes the paper by summarizing our result work and outlining future work.

## 2. Related work:

In this segment, we see some existing works related to opinion spam detection and the various different methods used by researchers to detect fake reviews. Jindal et al was the first who mentioned the opinion spam in Amazon reviews and provided some benchmark solution to detect them. There are two main categories of opinion spam detection research approach: review spam (textual) and review spammer (behavioural). Ott et al. developed a textual based detection approach in which they used n-gram analysis and term frequency to detect deceptive reviews. They collected a "gold-standard" dataset by gathering deceptive reviews of Chicago hotels from Amazon Mechanical Turk and honest reviews from TripAdvisor. They split all the reviews into positive and negative groups. Then they implemented n-gram analysis technique and term frequency for feature extraction technique to identify deceptive reviews. They achieved 86% accuracy using support vector machine (SVM).

Another textual based detection model was developed by Mukherjee et al. who used the Ott et al's dataset for fake reviews detection. Mukherjee et al. discussed in their research paper that the machine-generated reviews do not count as real world deceptive reviews as these do not represent opinion spams. They decided to test their model on Yelp dataset. Even they used Ott et al model to test the Yelp dataset and it achieved only 67.8%. Thus, they discussed that the results from models trained using machine generated fake reviews are not accurate and the methods are not useful for detecting real-world data. However, n-gram features are still useful for detecting deceptive reviews.

Fei et al. introduced a novel technique which is called burst detection mechanism and it was able to identify the customers who give deceptive review. Then they modeled their mechanism using markov random field and added a loopy belief propagation method to identify deceptive spammers in candidate bursts. They achieved 77.6% accuracy with the proposed approach.

Another related approach is introduced by Xie et al to detect singleton review (SR). According to their study, more than 90% of reviewers post only one review. The size of a singleton reviews is huge compared to size of the nonsingleton reviews. The author also explained that if the singleton reviews are continuously increasing within a rapid duration, then the fake reviewers are attempting to manipulate a product's reputation or rating.

Li et al. decided to focus on reviews that received high numbers of reviewer's votes and comments. The author assumed that reviews with low number of reviewer's votes are more suspicious than reviews with a high number of votes and they considered that it is more likely fake review. They investigated supervised machine learning techniques such as SVM, NB, and LR to identify review spam. They achieved 0.58 F-score using NB method which gives much better results compared to other methods that rely on behavior features.

Feng et al. discovered that there is a connection between the distribution anomaly and fake reviews detection. They assumed that some business agent used to hire spammers for writing fake reviews. For evaluation, they used an Ott et al "gold-standard dataset which contains 400 deceptive and truthful reviews. They achieved 72.5% accuracy on their test dataset. The author mentioned that the proposed method is very effective in detecting suspicious burst within a window of time. However, it is not an effective method for detecting whether users reviews are fake or truthful.

Mukherjee et al. developed an author spamicity model (ASM) to detect suspicious spammers based on their behavior track. They categorized the reviewers into two groups, spammers and nonspammers. Unlike previous research papers on behavioral analysis, they proposed a model for detecting deceptive reviews using an unsupervised Bayesian inference framework. The author results showcase that their proposed model is much more efficient and outperforms other supervised machine learning models.

## 3. Fake review detection model:

### 3.1 Approach overview:

The general approach for fake review detection starts with data pre-processing the dataset by removing unnecessary special characters, punctuation, stop words, and unnecessary words from the data. Then we've performed lemmatization for feature extraction in the cleaned dataset. The last step in the classification process is to train the classifier with the extracted features. We compared five different machine learning algorithms, namely, support vector machine (SVM), linear support vector machines (LSVM), passive aggressive classifier (PA), logistic regression (LR), and multinomial naive bayes (NB). Figure explains the work flow of the fake review detection model. The figure shows the proposed methodology of deceptive review detection.

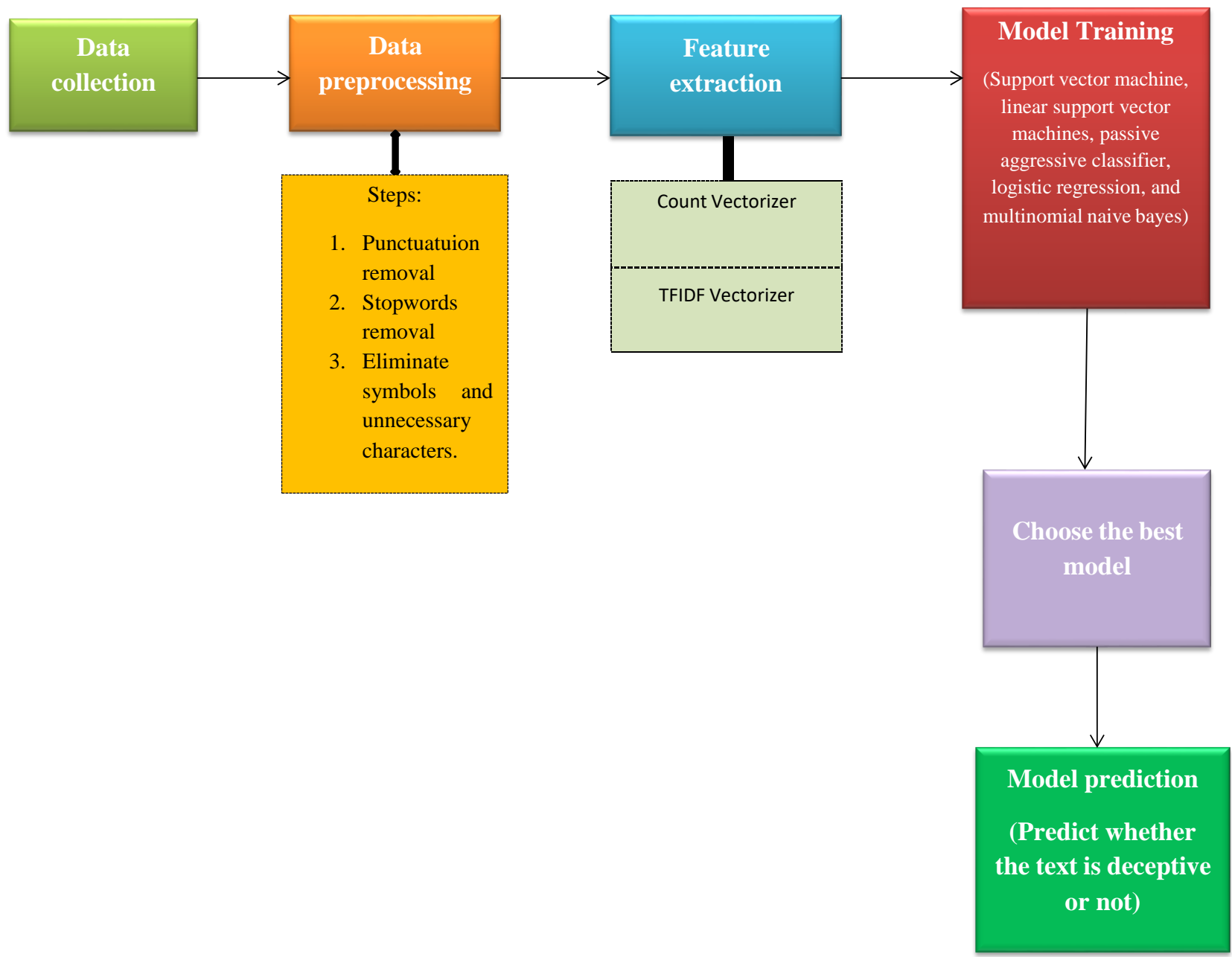

**Fig.1. Proposed Methodology of deceptive review detection**

### 3.2 Data Pre-processing:

In any machine learning task, data cleaning or pre-processing is the most important thing. When it comes to unstructured data, the process is much more important. Data pre-processing involves a lot of techniques include removing punctuations, removing URLs, removing stop words, lower casing, tokenization, stemming and lemmatization. After removing irrelevant information using preprocessing techniques, our data is now ready for feature extraction.

### 3.3 Tokenization:

Tokenization is one of the most important techniques that we use in natural language processing. It is a first and foremost step before applying any other natural language techniques. It is the process of dividing the given text data into smaller one called tokens. Alphanumeric characters, punctuation marks, and others special characters can be considered as tokens. The text is divided into individual words called tokens. For example, if we have a sentence like "the food is tasty", tokenization will split into the following tokens "the", "food", "is", "tasty".

### 3.4 Stop Words Removal:

Stop words are the common words which are used in day to day English language. Generally articles, conjunctions, interjections, prepositions, and some pronouns are considered as stop words. Common examples of the stop words are for, from, how, in, is, of, on, or, that, the, these, this, too, was, what, when, where, who, will, and so on. These words do not carry any useful meaning and are removed from each text document.

### 3.5 Lemmatization:

Lemmatization is the process of converting tokenized word into another form of word which is basically understand by human being. It reduces the inflectional forms of each word into a common root word. For example, the words "singing," "sang," and "singer" will be reduced to the word "sing." It takes a lot of time but it is very efficient and used in chatbot applications. So in this paper, we use lemmatization for data preprocessing.

### 3.6 Feature Extraction:

We cannot directly give the text to our machine learning model. We need to convert the words into numerical or vector form. Thus, it is best to perform feature extraction technique to reduce the text feature. We used two different features extraction methods, namely, count vectorizer or bag of words and term frequency inverse document frequency (TF-IDF). The methods are explained in the following section.

### 3.6.1 Count Vectorizer:

Using count vectorizer, we can convert variable-length texts into a fixed-length vector form and n-grams. In simple, we convert a text into its equivalent vector of numbers by using bag-of-words (BoW) technique. A bag-of-words is defined as the description of text format that describes the event of words within a text document. Count vectorizer generates its own matrix in which each word is constituted by a column of the matrix, and each sample text from the sentence is a row in the matrix.

### 3.6.2 Term frequency inverse document frequency:

The Bag of Words (BoW) method is simple and works well, but it treats every word equally. There is some disadvantage in the Bag of Words. In Bag of words (Bow), some words have same representation and same weightage and so there is no semantics. In order to prevent this, we will use TF-IDF. Term frequency inverse document frequency (TF-IDF) is one of the method for feature extraction techniques in natural language processing. It measures how important a word is within a document relative to corpus. TF-IDF transforms a word to vector form by multiplying the term frequency (TF) with the inverse document frequency (IDF).

$$TF\text{-}IDF = TF * IDF$$

Term Frequency: It is the number of times the word appears in a document compared to the total number of words in the document.

$$TF = \frac{\text{number of repetitive words in the document}}{\text{total number of words in the document}}$$

Inverse Document Frequency: It is the total number of documents in the corpus compared to the number of documents in the corpus that containing the words with taking logarithmic calulations.

$$IDF = \log(\frac{\text{total number of the documents in the corpus}}{\text{number of documents in the corpus containing words}})$$

### 3.7 Implementation:

After extracting the features using either count vectorizer or TF-IDF, we train a machine learning model to decide whether a review is truthful or fake. Before applying extracting features to the model, we split the dataset into training and test sets using both train test split and K-fold cross validation (K=5). After that, we will implement five different classifiers to predict the class of the reviews, including support vector machine (SVM), linear support vector machines (LSVM), passive aggressive classifier (PA), logistic regression (LR), and multinomial naive bayes (NB).

## 4 Experimental Evaluations

In this section, we evaluated our proposed approach on Ott et al's dataset and discussed about obtained results.

### 4.1 Experiments overview:

We used a public dataset, gathered by Ott et al. The dataset is available in kaggle in the title "Deceptive opinion spam corpus". The dataset contains 1600 reviews which are classified into 800 truthful reviews and 800 fake reviews. The reviews are based on the twenty topmost hotels in Chicago. The dataset is collected from TripAdvisor and Amazon Mechanical Turk. The following information is available in the dataset.

- Review label (Fake or truthful)
- Hotel name
- Review sentiment or Polarity (Positive or negative)
- Review source (Trip Advisor, Mechanical Turk)
- Review text

We have taken review label and review text in the dataset for data pre-processing and we ignored the remaining ones. After pre-processing our dataset, we convert the string data into vector form using both count vectorizer and TF-IDF vectorizer. For feature extraction, n gram range and max feature play a prominent role for getting higher accuracy. We also noticed that increasing n gram number influences the overall accuracy of the model. We varied the max features ranging from 1000 to 50000 and changed n gram range from n=1, 2, 3, 4. At last n gram range = (1, 3) and max features = 11000 give us better accuracy compared to other features. Then we give the vector data to the machine learning models. Table 1 explains the accuracy of all machine learning models using count vectorizer with train test split.

**Table 1 : Accuracy of five machine learning models using count vectorizer with train test split**

| Classifier | Feature | Accuracy |
|---|---|---|
| Logistic Regression | BOW | 90.3% |
| Linear Support vector Machine | BOW | **91.8%** |
| Passive Aggressive Classifier | BOW | 90% |
| Multinomial Naïve Bayes | BOW | 89.3% |
| Support Vector Machine | BOW | 90.6% |

Table 1 explains the accuracy of five machine learning models using count vectorizer. We split the dataset into train and test set using train test split with test size = 0.2. It is found that the linear support vector machine classifier gives the highest accuracy of 91.8% using count vectorizer in bigram model. Support vector machine (SVM) and logistic regression classifiers have relatively close accuracy to LSVM. The lowest accuracy of 89.3% was achieved using multinomial naïve bayes (NB). In order to perform the different kinds of splitting the dataset, we take into consideration the k fold cross validation. We use 5 fold cross validation for splitting the dataset. In each validation splitting round, the dataset divided into 80% for training and 20% for testing.

**Table 2 : Accuracy of five machine learning model using count vectorizer with K fold cross validation**

| Classifier | Feature | Accuracy |
|---|---|---|
| Logistic Regression | BOW | 88.8% |
| Linear Support vector Machine | BOW | 87.8% |
| Passive Aggressive Classifier | BOW | 88.3% |
| Multinomial Naïve Bayes | BOW | 88.4% |
| Support Vector Machine | BOW | 86.2% |

Table 2 explains the accuracy of five machine learning models using count vectorizer with k fold cross validation. We split the dataset into train and test set using k fold cross validation with k = 5 (which means k = 1/5 = 0.2). For each folding, the data is equally split into 80:20 ratios. Using logistic regression, we achieved 88.8% accuracy compared to the other algorithms. It is noted that the passive aggressive classifier and multinomial naïve bayes have very close accuracy to logistic regression. So all the three models performed well on k fold cross validation. Support vector machine (SVM) and logistic regression classifiers have relatively close accuracy to LSVM. Support vector machine (SVM) gives us lowest accuracy compared to the other algorithms. Figure 1 shows the overall performance of all the five classifiers using count vectorizer with k fold and train test split. The figure clearly explains that the LSVM outperforms all the other classifiers in terms of accuracy achieving 91.8% accuracy. Using 5 fold cross validation, logistic regression performs well compared to other classifiers.

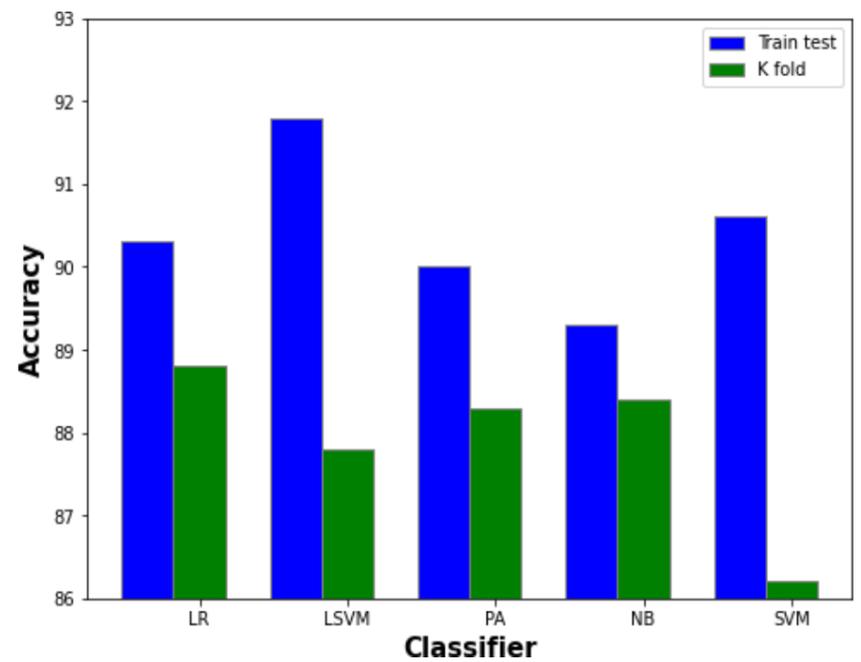

**Figure 2 : Overall performance of all the classifiers using count vectorizer**

**Table 3 : Accuracy of five machine learning models using TF-IDF vectorizer with train test split**

| Classifier | Feature | Accuracy |
|---|---|---|
| Logistic Regression | TF-IDF | 88.7% |
| Linear Support vector Machine | TF-IDF | 90.9% |
| Passive Aggressive Classifier | TF-IDF | **92.5%** |
| Multinomial Naïve Bayes | TF-IDF | 88.7% |
| Support Vector Machine | TF-IDF | 89% |

Table 3 summarizes the accuracy of five machine learning models using TF-IDF vectorizer. As we mentioned before, we tested our models using both train test split and k fold cross validation. So the passive aggressive classifier gives the highest accuracy of 92.5% using TF-IDF vectorizer in bigram model. Linear support vector machine (SVM) classifier also performed well and achieved above 90%. Logistic regression and multinomial naïve bayes (NB) give us little bit less accuracy compared to other classifiers.

**Table 4 : Accuracy of five machine learning model using TF-IDF vectorizer with K fold cross validation**

| Classifier | Feature | Accuracy |
|---|---|---|
| Logistic Regression | TF-IDF | 89.1% |
| Linear Support vector Machine | TF-IDF | 89.2% |
| Passive Aggressive Classifier | TF-IDF | 89.4% |
| Multinomial Naïve Bayes | TF-IDF | 88.1% |
| Support Vector Machine | TF-IDF | 89.1% |

Table 4 summarizes the accuracy of five machine learning models using TF-IDF vectorizer with k fold cross validation. Using passive aggressive classifier, we achieved a little bit 89.4% accuracy compared to the other algorithms. It is found that the linear support vector machine (LSVM), support vector machine (SVM) and logistic regression classifiers have bit close to passive aggressive classifier accuracy. Finally we have observed that the passive aggressive classifier performed well on this dataset and gave us better accuracy both in train test split and k fold cross validation using TF-IDF vectorizer. Similar to the results obtained in figure 1, linear-support vector machine (LSVM) achieved better results in TF-IDF. However, the highest accuracy was achieved using passive aggressive classifier, 92.5%. This classifier performs well on TF-IDF vectorizer with both train test split and k fold cross validation. Figure 2 illustrates the overall performance of all the classifiers using TF-IDF vectorizer in both train test split and k fold cross validation. The figure clearly explains that the passive aggressive outperforms all the other classifiers in terms of accuracy achieving 92.5% accuracy. It also performs well on both train test split and k fold cross validation using TF-IDF vectorizer.

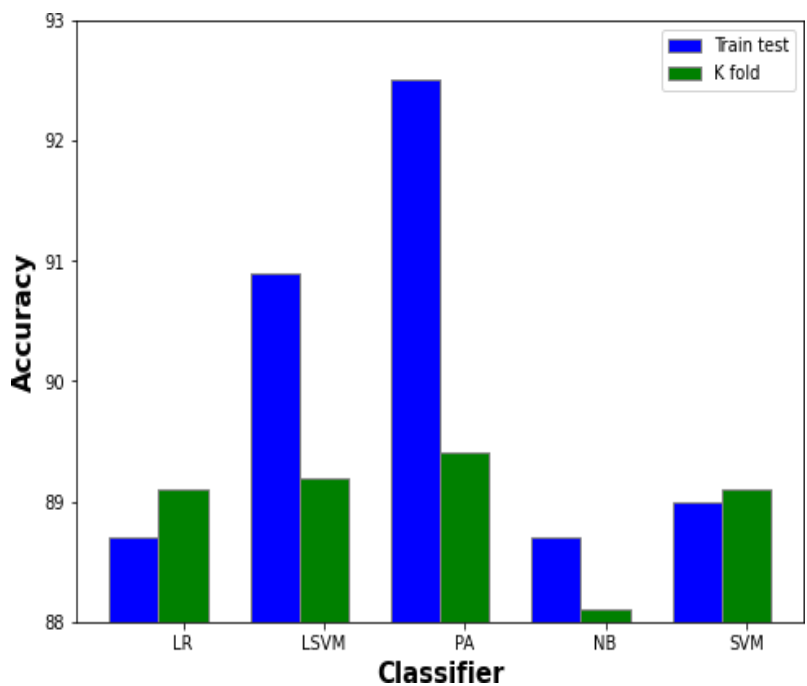

Figure 3 : Overall performances of all the classifiers using TF-IDF vectorizer

### 4.2 Performance Comparison and discussion:

As mentioned earlier, our model gives the best result compared to the existing work. We investigated two feature extraction techniques and our model outperforms the existing approaches. Comparison of previous work and our model is shown in the table 5. For detecting deceptive review, we have compared our model to Ahmed et al's "Detecting opinion spams and fake news using text classification".

In the research paper, they used stemming for data preprocessing in the dataset. Ahmed et al. developed a fake review detection model that binds text analysis using n-gram features and TF-IDF metrics. They achieved 90% accuracy using LSVM, which is slightly higher than the 89% achieved by Ott et al. on the same dataset.

Using the same dataset, we used lemmatization for data cleaning. Then we investigated both feature extraction techniques (count vectorizer and TFIDF) and analysed that max features and n gram range in feature extraction techniques play an important role for getting higher accuracy. Then we've selected max_features = 11000 and n_gram_range = (1, 3) for getting higher accuracy after running different ranges. Our proposed model achieved 91.8% accuracy for using linear support vector machine with count vectorizer and 92.5% accuracy for using passive aggressive classifier with TF-IDF vectorizer which is slightly higher than Ahmed et al's results.

Table 5 : Performance comparison of our model with existing works

| Machine Learning model | Feature extraction techniques | Accuracy | Research |
|---|---|---|---|
| Support vector Machine | LIWC +Bigram | 89% | Ott et al. result |
| Linear Support vector Machine | TFIDF Vectorizer | 90% | Ahmed et al. results |
| Linear Support vector Machine | Count Vectorizer | 91.8% | Our results |
| Passive Aggressive Classifier | TFIDF Vectorizer | 92.5% | Our results |

### 4.3 Applying Deep Learning technique:

#### 4.3.1 Data augmentation:

Table 6 shows the accuracy of deep learning models. Also, the accuracy is very moderate compared to machine learning model classifier accuracy. Since the dataset is too small, it is not good enough to train any deep learning models. After cleaning the data, we split the data into 80:20 ratios. We've performed data augmentation technique. So we've augmented the data with synonym replacement technique. Each sentence is augmented to ten times (n=10) and finally, we've created augmented data. The original dataset is 1600 and now the augmented dataset is 14400. We didn't perform any data augmentation technique on test data (20%) and we performed augmentation only on training data (80%). The library we've used for data augmentation is nlpaug.augmenter.word and aug_src is wordnet parameter.

Table 6 : Accuracy of deep learning models in deceptive text classification

| Deep Learning model | Library | Training set accuracy | Test set accuracy |
|---|---|---|---|
| LSTM | Word2Vec+Gensim | 98% | 85.9% |
| LSTM | GloVe | 99.6% | 81.56% |
| BERT (distilbert) | Hugging Face | 95.7% | 85.6% |
| RoBERTa | Fastai | 97.3% | 90.9% |
| Simple CNN | SpaCy | 100% | 89% |

Table 7 : Accuracy of deep learning models with augmentation in deceptive text classification

| Deep Learning model | Data Augmentation | Training set accuracy | Test set accuracy |
|---|---|---|---|
| RoBERTa | 80% training data (data augmentation) + 20% test data (no augmentation) | 100% | 99.68% |
| RoBERTa | Entire data augmentation | 100% | **99.8%** |

Table 7 explains the augmentation results of deceptive review detection. Firstly we pre-process the augmented data by removing punctuations and stop words. Then we convert the augmented data into word vectors by Word2Vec method in gensim library and applied tokenization to the vector data. Then we give the vector data to the LSTM model. Using LSTM model, we achieved 98% accuracy in training data and 85.6% accuracy in test data. Secondly, we use GloVe model. We've used Stanford pre-trained embedding vector for a dictionary and applied tokenization to the dataset which will create word vectors. Then we give this vector data to the LSTM model. Then we got 81.56% in test data and 99% in training data. We also implements transformers models like BERT, RoBERTa. We achieved 86% accuracy using BERT model and 91% accuracy using RoBERTa model. After applying K-fold cross validation (k=5), we achieved 99.5% accuracy in RoBERTa model.

## 5. Conclusion:

In recent years, the opinion spam problems become a growing research area due to the plenty of online-generated content. Even fake reviews play a prominent role in ecommerce industry and lot of other social media platforms. Now-a-days, anyone can post a fake review on online websites. Many ecommerce businesses mislead their customers by posting good reviews on the particular product. It is strenuous for customers to identify good products from defective ones based on reviews.

So to avoid this, our research paper focused on detecting fake review using different feature extraction techniques. Firstly, we have investigated different feature extraction techniques used by many researchers. Then, we outlined some traditional machine learning approaches for deceptive review detection with summary tables and charts. We also provided a comparative analysis of some existing work and our proposed work in fake review detection. The outcomes showed that passive aggressive classifier achieved the highest accuracy on the Ott el al's dataset which means our results shows 2.78% greater than Ahmed et al results.

Usually supervised machine learning uses a labelled dataset to predict whether the review is deceptive or not, which is hard to obtain in real world fake review detection. So we can conclude that most of the existing approaches related to supervised machine learning models to detect fake reviews. Thus, we would like to explore more on deep learning neural network to detect deceptive review and tackle different ways to solve this problem. In addition, we would also like to add feature selection technique like chi square for this fake review detection problem.

**References:**

1. Ahmed H, Traore I, Saad S. Detecting opinion spams and fake news using text classification, Security and Privacy, 2018;1:e9. https://doi.org/10.1001/spy2.

2. Li J, Ott M, Cardie C, Hovy E. 2014. Towards a general rule for identifying deceptive opinion spam. Paper presented at: Proceedings of the 52nd Annual Meeting of the Association for Computational Linguistics; June 23-25, 2014:1566–1576; Baltimore, MD: ACL

3. Mukherjee A, Liu B, Glance N. Spotting fake reviewer groups in consumer reviews. Paper presented at: Proceedings of the 21st international conference on World Wide Web; 2012. Lyon, France: ACM.

4. Jindal N, Liu B. Opinion spam and analysis. Proceedings of the 2008 International Conference on Web Search and Data Mining. New York, NY: ACM; 2008:219-230. https://doi.org/10.1145/1341531.1341560

5. Jindal, N., & Liu, B. Analyzing and detecting review spam. In Proceedings of 7th IEEE International Conference on Data Mining (ICDM), pp. 547–552. https://doi.org/10.1109/icdm.2007.6

6. Fei, G., Mukherjee, A., Liu, B., Hsu, M., Castellanos, M., & Ghosh, R. (2013). Exploiting bursti-ness in reviews for review spammer detection. In Proceedings of 7th International AAAI Conference on Weblogs and Social Media (ICWSM), pp. 175–184.

7. Feng, S., Banerjee, R., & Choi, Y. (2012). Syntactic stylometry for deception detection. In Proceed-ings of the 50th Annual Meeting of the Association for Computational Linguistics (Volume 2: Short Papers), pp. 171–175

8. Jindal, N., & Liu, B. (2007b). Review spam detection. In Proceedings of the 16th International Conference on World Wide Web (WWW), pp. 1189–1190. https://doi.org/10.1145/1242572.1242759.

9. Li, F., Huang, M., Yang, Y., & Zhu, X. Learning to identify review spam. In Proceeding of the 22nd International Joint Conference on Artificial Intelligence (IJCAI'11), pp. 2488–2493.

10. Li, J., Cardie, C., & Li, S. Topicspam: a topic-model based approach for spam detection. In Proceedings of the 51st Annual Meeting of the Association for Computational Linguistics (Volume 2: Short Papers), pp. 217–221.

11. Li, J., Ott, M., Cardie, C., & Hovy, E. Towards a general rule for identifying deceptive opinion spam. In Proceedings of the 52nd Annual Meeting of the Association for Computational Linguistics (Volume 1: Long Papers), pp. 1566–1576.

12. Ott, M., Choi, Y., Cardie, C., & Hancock, J.T. Finding deceptive opinion spam by any stretch of the imagination. In Proceedings of the 49th Annual Meeting of the Association for Computational Linguistics: Human Language Technologies-Volume 1, pp. 309–319.

13. Xie, S., Wang, G., Lin, S., & Yu, P.S. Review spam detection via temporal pattern discovery. In Proceedings of the 18th ACM SIGKDD International Conference on Knowledge Discovery and Data Mining, pp. 823–831. https://doi.org/10.1145/2339530.2339662.